\documentclass[aps,epsf,preprint]{revtex4}
\usepackage{amssymb}

\def\1{{\bf 1}}
\def\[{\left[}
\def\]{\right]}
\def\be{\begin{eqnarray}}
\def\ee{\end{eqnarray}}
\def\bm{\begin{matrix}}
\def\em{\end{matrix}}
\def\nn{\nonumber}
\def\({\left(}
\def\){\right)}

\def\eq#1{(\ref{#1})}

\def\o{\omega}

\def\f{\phi}

\def\G{{\cal G}}
\def\C{{\cal C}}

\def\x{\times}

\def\labels#1{\quad [#1]\label{#1}}
\def\edc{\end{document}}

\def\bn{\begin{enumerate}}

\def\en{\end{enumerate}}

\def\g{\gamma}

\def\rt{\sqrt{3}}

\def\diag{{\rm diag}}

\def\ba{\begin{array}}
\def\ea{\end{array}}
\def\bc{\begin{center}}
\def\ec{\end{center}}

\def\edoc{\end{document}}
\def\nl{\newline}
\def\hs{\hskip.5cm}
\def\ni{\noindent}
\def\^{$\wedge$}
\begin{document}

\title{Leptonic Mixing and Group Structure Constants}
\author{C.S. Lam}
\address{Department of Physics, McGill University\\
 Montreal, Q.C., Canada H3A 2T8\\
and\\
Department of Physics and Astronomy, University of British Columbia,  Vancouver, BC, Canada V6T 1Z1 \\
Email: Lam@physics.mcgill.ca}

\begin{abstract}
Hernandez and Smirnov  discovered an interesting formula  to parametrize each column of a neutrino mixing matrix
by six integers related to the residual symmetry. We point out that these six integers are not independent,
and propose a way to find the allowed combinations using structure constants of finite groups.

\end{abstract}
\narrowtext
\maketitle

\section{Introduction}
Symmetry of leptonic mixing can be studied from the top down, or from the bottom up. In the former case, a symmetry group
$\G$ is introduced from which the leptonic mixing patterns are computed. In the latter case, $\G$ is derived
from the residual symmetries compatible with experimental mixing. In this article we discuss the latter.

It was recently found \cite{HS,Hu} that each column of a neutrino mixing matrix can be parametrized by
six integers related to residual symmetry. This is a great step forward because this formula for mixing vectors
is explicit and easily calculable. 
In contrast, it is considerably harder and much more lengthy to produce mixing vectors from the top
down, because in that case many eigenvectors of many group elements of many 3-dimensional irreducible representations of many
groups have to be computed \cite{Lam1, HLL}.

Nevertheless, there are hidden difficulties with this bottom-up apprach.
 As we shall see in Sec.~II, without a constraint between these six parameters, there may be no group containing the intended residual
symmetry, much less a finite one with a moderate order. The purpose of this article is to find out 
the constraints between these parameters.

After a review of the main formula \cite{HS} in Sec.~II, and some of its consequences,
we will embark on a way to search  for the constraints in Sec.~III.
We need to do a scan  to find the allowed parameters, but this scan is much simpler than the
top-down scan \cite{Lam1, HLL} because 
only class functions are needed here rather than the individual group elements. Examples of the search are given in Sec.~IV,
with the help of two GAP-language \cite{GAP} programs discussed in Appendix A.

\section{Review of the Parametrization}
We follow the notations of \cite{HS}. The symmetry generator of $Z_m$ in the charged-lepton sector is
 denoted by $T$, rather than $F$ as in \cite{Lam1}, and those for the $Z_2$ or $Z_2\x Z_2$ symmetry 
in the neutrino sector are denoted by $S_{iU}$,
rather than $G_i$. Following \cite{HS}, $\det(T)=\det(S_{iU})=1$ is assumed, and the order of $W_{iU}:=S_{iU}T$
is taken to be $p$. Most of time we will discuss a partial $Z_2$ symmetry and will drop the subscript $i$ in $S_U$
and $W_U$.

The absolute value of a column of the mixing matrix $U$ can be parametrized by six integers $(m,k_e,k_\mu,p, s_e, s_\mu)$
to be \cite{HS}
\be 
|U_{ei}|^2&=&{a_R\cos{\f_e\over 2}+\cos{3\f_e\over 2}-a_I\sin{\f_e\over 2}\over 4\sin{\f_{e\mu}\over2}\sin{\f_{\tau e}\over 2}}\ ,\nn\\
|U_{\mu i}|^2&=&{a_R\cos{\f_\mu\over 2}+\cos{3\f_\mu\over 2}-a_I\sin{\f_\mu\over 2}\over 4\sin{\f_{e\mu}\over2}\sin{\f_{\mu\tau }\over 2}}\ ,\nn\\
|U_{\tau i}|^2&=&{a_R\cos{\f_\tau\over 2}+\cos{3\f_\tau\over 2}-a_I\sin{\f_\tau\over 2}\over 4\sin{\f_{\tau e}\over2}\sin{\f_{\mu\tau }\over 2}}\ ,\labels{mv}
\ee
where  $i$ labels the column. The integer parameters $(m,k_e,k_\mu)$
are defined via the $Z_m$ generator $T=\diag(\o_m^{k_e}, \o_m^{k_\mu}, \o_m^{k_\tau})$,
where $\o_m:=\exp(2\pi i/m)$. The angles are  $\f_e:=2\pi  k_e/m, \ \f_\mu:=2\pi k_\mu/m, \ \f_\tau:=2\pi  k_\tau/m,\ 
\f_{e\mu}:=\f_e-\f_\mu,\ \f_{\mu\tau}:=\f_\mu-\f_\tau$, and $\f_{\tau e}:=\f_\tau-\f_e$. The parameter $k_\tau=-(k_e+k_\mu)$
is determined by $\det(T)=1$.  The quantities
$a_R$ and $a_I$ are the real and imaginary parts of  $a:={\rm Tr}(W_U)=\o_p^{s_e}+\o_p^{s_\mu}+\o_p^{s_\tau}$, where
 $\o_p^{s_e},\ \o_p^{s_\mu}, \o_p^{s_\tau}$ are the eigenvalues of $W_U=S_UT$.  Since $\det(W_U)=1$, it follows that
$s_e+s_\mu+s_\tau=0$, so \eq{mv} is parametrized by six integers $(m, k_e, k_\mu, p, s_e, s_\mu)$.
The range of these integers are
$m\ge 3$, $m>k_e, k_\mu\ge 0$, and $p>s_e, s_\mu\ge 0$, but they cannot be chosen independently.

One indication that the six integers are not arbitrary is the following.  The right-hand side of \eq{mv} is
 not guaranteed to be
non-negative but the left-hand side must be. For example, with $(m,p)=(3,5)$, there is at least one negative
entry in \eq{mv} if we take  $(k_e,k_\mu,s_e,s_\mu)$ to be either (0,1,0,0), (0,1,1,1), (0,1,1,2),
(0,1,1,3), (0,1,2,2),  or (0,1,2,4), so these choices of parameters are forbidden.

The reason for this failure is that a group structure cannot emerge from these choices, as we shall see in Sec.~IVB.

Many known examples in the literature relies on $m=3$ and 
$T=\diag(1,\o,\o^2)$. In fact, this $T$ is unique up to permutation for $m=3$ if it is to be non-degenerate. 
For that $T$, \eq{mv} is reduced to
\be 
|U_{ei}|^2&=&{1\over 3}(a_R+1),\nn\\
|U_{\mu i}|^2&=&{1\over 6}(-a_R+2+\rt a_I),\nn\\
|U_{\tau i}|^2&=&{1\over 6}(-a_R+2-\rt a_I),\labels{mv3}\ee
so the  equivalent mixing vectors are completely controlled by $a={\rm Tr}(W_U)$. We call two mixing vectors  {\it equivalent} if they differ only by a permutation of their components.

Table 1 shows some known cases  taken from \cite{HS, Hu, Lam1, HLL, TFH, Lam2} with this $T$.
Here the parameters $(m, k_e, k_\mu)=(3,0,1)$ are fixed. The first column in the Table gives the GAP-designation 
of the group in its Small Group library \cite{GAP},
the second column the common name of the group, the third column
the mixing vector $|v|=(|U_{ei}|, |U_{\mu i}|, |U_{\tau i}|)$, or one of
its permutations. The last two columns give $(p,s_e,s_\mu)$ and the corresponding $(a_R, a_I)$ parameters that
give rise to this $|v|$. \eq{mv3} can be used to check that the listed $|v|$ indeed corresponds to the listed $(a_R, a_I)$.

$$\ba{|c|c|c|c|c|}\hline
{\rm SmallGroup}&{\rm Group}&|v|&(p,s_e,s_\mu)&(a_R,a_I)\\ \hline
\[12,3\]&A_4&[.577,.577,.577]&(3,0,1)&(0,0)\\
\[96,64\]&\Delta(96)&[.577, .789, .211]&(8,1,2)& (0, 1)\\
\[24,12\]&S_4&[.816,.408,.408]&(4,0,1)&(1,0)\\
\[150,5\]&\Delta(150)&[.480, .812, .332]&(10,1,3)&(-.309, .951)\\
&&[.777, .607, .170],&(10,1,2)&(.809, .588)\\
\[162,14\]&&[.525, .279, .804]& (18,7,13)&(-.174, -.985)\\
\[168,42\]&PSL(2,Z_7)&[.408, .894, .187]&(7,1,2)& (-.500, 1.32)\\
\[294,7\]&\Delta(294)&[.354, .814, .460]&(14,1,5)&(-.623, .782)\\
\[384,568\]&\Delta(384)&[.312, .810, .497)&(16,1,6)& (-.707, .707)\\
\[486,61\]&\Delta(486)&[.525, .279, .804]& (18,7,13)&(-.174, -.985)\\
\[600,179\]&\Delta(600)&[.480, .812, .332]&(10,1,3)&(-.309, .951)\\
\hline\ea$$
\bc Table 1. Some mixing vectors obtained from $(m,k_e.k_\mu)=(3,0,1)$\ec

\section{Group Structure}
Suppose $\G$ is a finite group that contains the elements $T, S_U, W_U=S_UT$, of orders $m, 2, p$ respectively. Then its order $|\G|$ must be an integer multiple of the least common multiple of 2, $m$, and $p$.
Moreover, the product of an order-$2$ element $S_U$ and an order-$m$ element $T$ must produce an order-$p$
element $W_U$. Whether this can be realized or not depends on $\G$, and  is determined by the {\it structure
constant} of that group.

\subsection{Structure Constant}
Let $\C_i$ represent the $i$th conjugacy class of $\G$, and $\C_i\C_j$ the set of all products $xy$ with $x\in\C_i$
and $y\in\C_j$. It is not hard to see that $\C_i\C_j$ must cover all classes $\C_k$ an integer number of times so that
we can write
\be \C_i\C_j=\sum_k\g_{ijk}\C_k.\labels{stcon}\ee
The non-negative integers $\g_{ijk}$ are known as  {\it structure constants} of the finite group $\G$. 
It is symmetric in $i$ and $j$ because $x\C_jx^{-1}=\C_j$, hence $\C_i\C_j=\C_j\C_i$.
They can be
calculated from the characters $\chi_i^\mu$ of class $\C_i$ and irreducible representation $\mu$ by the formula
\be \g_{ijk}={|\C_i|\ |C_j|\over |\G|}\sum_\mu {1\over n_\mu}\chi^\mu_i\chi^\mu_j\chi^\mu_k,\labels{gabcchi}\ee
where $|\C_i|$ is the size of class $\C_i$ and $n_\mu$ is the dimension of the irreducible representation $\mu$.
Hence $|\C_k|\g_{ijk}$ is completely symmetric in $i,j,k$.
 
Let $c_o$ denote the list of classes whose elements have order $o$. Instead of writing the list as $\{\C_i, \C_j, \cdots\}$,
we simply write it as $c_o=\{i, j, \cdots\}$.
Given a choice of $(m,p)$, we need to find the groups $\G$ that are `structurally sound', in the sense that
there is at least one $\g_{ijk}>0$ with $i\in c_2,\ j\in c_m$, and $k\in c_p$. A program in GAP language to search
for such groups is presented in Appendix A.

\subsection{Eigenvalues}
Given a group $\G$ that is structurally sound for the parameters $(m,p)$, the next task is to find out the allowed
choice of  $(k_e, k_\mu, s_e, s_\mu)$ for this group. 
Since these parameters are related to the eigenvalues of $T$ and 
$W_U$, and all elements within a class have the same eigenvalues, they can be obtained by searching  the eigenvalues
of classes $\C_j\in c_m$ and $\C_k\in c_p$ for which $\g_{ijk}>0$. There could be different sets of 
eigenvalues for different 3-dimensional irreducible representations IR$_3$ and all of them should be found.
For $c_p$, it is actually only the trace  that is needed, not the individual eigenvalues. For a 3-dimensional
irreducible representation $\mu$, $a$ is simply the character $\chi_j^\mu$.  
The output of the program
in Appendix A contains these allowed eigenvalues.

\section{Examples}
Let us illustrate the discussion of Sec.~III with  some examples.

\subsection{$(m, k_e, k_\mu)=(3, 0, 1),\quad p=10$} 
For $(m,p)=(3,10)$, the least common multiple of (2, 3, 10) is 30, so we need to consider only groups $\G$ whose order is
a multiple of 30. With $\det(T)=\det(S_U)=\det(W_U)=1$,
$\G$ can be confined to finite subgroups of $SU(3)$, and we will limit ourselves to those  with an order $<512$. From the list of such subgroups \cite{Ludl},
only three groups have the right order. They are the Small Groups [60,5] =$A_5$, [150,5]=$\Delta(150)$, and [300,43]=$\Delta(300)$.

Table 2 shows these groups, as well as  the allowed parameters $(s_e, s_\mu)$ and $a=a_r+ia_I$
when the group is structurally sound.  This table is taken from the output of
the programs `gpsearch'  and `gpsu3' in Appendix A.

The last four columns in the Table 
display the order-$o$ classes $c_o$, for $o=2,\ o=m=3,\ o=p=10$, and  the list IR$_3$ of  3-dimensional
irreducible representations of the group. We see that [60,5] is not structurally sound because it lacks order-10 elements.
The group [300, 43] does have plenty of order-10 elements, but it fails because there is no non-vanishing
structure constants between classes of orders 2, 3, and 10. That leaves only the group [150,5].

$\Delta(150)=[150,5]$ has one class each of order 2 and 3, four classes of order 10, and eight 3-dimensional irreducible
representations. However, only the combinations in column 4 have non-zero structure constants and unit determinants.
These combinations are consistent with those listed in Table 6 of \cite{Lam2}, though they are obtained in completely
different ways. The allowed parameters given in columns 2 and 3 form two complex-conjugated pairs, with 
both $(s_e,s_\mu)=(1,2)$ and $(s_e,s_\mu)=(1,3)$ appeared in Table 1. According to \eq{mv3}, exchanging
$a$ with $a^*$ only permutes $|v_2|$ and $|v_3|$, so the mixing vectors of $(s_e,s_\mu)=(3,8)$ and (1,2) are equivalent,
and those of (4,7) and (1,3) are equivalent. 

Any pair $(s_e,s_\mu)$ in the Table can be replaced by an equivalent pair $(s_\alpha, s_\beta)$ in which $\alpha, \beta$
are chosen from $e, \mu, \tau$, because they all give rise to the same $a$. For example, take the pair $(s_e,s_\mu)=(1,2)$.
Since $s_e+s_\mu+s_\tau=0$ mod $p$, it follows that
 $s_\tau=7$, therefore any pair taken from 1,2,7 will give the same $a$ and the same
mixing vector.

In principle, both $s_e$ and $s_\mu$ can vary from 0 to 9, but according to
Table 2, no combination $(s_e, s_\mu)$ other than these four (and their equivalent ones as explained
in the last paragraph) are allowed, at least not for a group of order $<512$.

$$\ba{|c|c|c|c|c|c|c|c|}\hline
{\rm SmallGroup}&(s_e, s_\mu)&a=(a_R,a_I)&(i,j,k,{\rm IR}_3)&c_2&c_3&c_{10}&{\rm IR}_3\\ \hline
[60,5]&-&-&-&2&3&-&2,3\\ \hline
[150,5]&(1,2)&\o_{10}=(.809, .588)&(2,3,7,9),(2,3,7,11)&2&3&6,7,10,11&4\!\!-\!\!11\\
&&&(2,3,10,7),(2,3,11,11)&&&&\\
&(1,3)&\o_{10}^3=(-.309, .951)&(2,3,6,9),(2,3,10,5)&&&&\\
&&&(2,3,11,7)&&&&\\
&(4,7)&\o_{10}^7=(-.309,-.951)&(2,3,6,7),(2,3,7,5)&&&&\\
&&&(2,3,10,11)&&&&\\
&(3,8)&\o_{10}^9=(.809, -.588)&(2,3,6,11),(2,3,7,7)&&&&\\
&&&(2,3,10,9)&&&&\\ \hline
[300,43]&-&-&-&3&2,6&7\!\!-\!\!10, 13\!\!-\!\!18& 4\!\!-\!\!36\\
&&&&&&22\!\!-\!\!29, 31\!\!-\!\!36&\\
\hline\ea$$
\bc Table 2. Allowed groups and allowed parameters for $(m,p)=(3,10)$\ec

\subsection{$(m, k_e, k_\mu)=(3, 0, 1),\quad p=5$} 
Since the least common multiple of (2, 3, 5) is also 30, the groups to be considered here are identical to those in Sec.~IVA,
namely, [60,5], [150,5], and [300,43]. Table 3 is the equivalent of Table 2, now for $p=5$. We see that 
[60,5]=$A_5$ is the only structurally sound group for $(m,p)=(3,5)$.

In principle, $s_e$ and $s_\mu$ can range from 0 to 4, but we saw in Sec.~II that the combinations
$(s_e,s_\mu)=(0,0), (1,2), (1,3), (2,2), (2,4)$ cannot be allowed because they give rise to a negative $|U_{\alpha i}|^2$.
Let us see why this is so from the group theoretical point of view.
According to Table 3, the allowed combinations of $(s_e,s_\mu)$ are (1,4) and (2,3), both leading to $s_\tau=0$, 
so the allowed pairs should be chosen either from 1,4,0, or 2,3,0, but none of the forbidden pairs in Sec.~II are one of these,
which is why none are allowed.

$$\ba{|c|c|c|c|c|c|c|c|}\hline
{\rm SmallGroup}&(s_e, s_\mu)&a=(a_R,a_I)&(i,j,k,{\rm IR}_3)&c_2&c_3&c_{5}&{\rm IR}_3\\ \hline
[60,5]&(1,4)&-\o_5^2(1+\o_5)=(1.618,0)&(2,3,5,2),(2,3,4,3)&2&3&4,5&2,3\\ 
&(2,3)&-\o_5(1+\o_5^3)=(-0.615,0)&(2,3,4,2),(2,3,5,3)&&&&\\
\hline
[150,5]&-&-&-&2&3&4,5,8&4\!\!-\!\!11\\ 
&&&&&&9,12,13&\\ \hline
[300,43]&-&-&-&3&2,6&4,5,11& 4\!\!-\!\!36\\
&&&&&&12,19\!\!-\!\!21,30&\\
\hline\ea$$
\bc Table 3. Allowed groups and allowed parameters for $(m,p)=(3,5)$\ec

\subsection{$(m,p)=(4,7)$}
The programs `gpsearch' and `gpsu3' in Appendix  A can be used on any $(m,p)$, not just $m=3$. To illustrate that, consider the example $(m,p)=(4,7)$.
In this case, the least common multiple of $(2,m,p)$ is 28. Among finite $SU(3)$ subgroups of order $<512$, only
the groups [84,11], [168,42], [336,57] have orders which are integer multiples of 28.  Of these three groups, only the
group $[168,42]=\Sigma(168)=PSL(3,2)=PSL(2,7)$ is structurally sound, as seen in Table 4.
The allowed parameters $(s_e,s_\mu)=(1,2)$ with $s_\tau=4$, and $(s_e,s_\mu)=(3,5)$ with $s_\tau=6$,  
are complex conjugate of each other. The values of $(k_e, k_\mu, k_\tau)$  not listed in the Table are both (0,1,3).
These assignments are those in known examples \cite{HS, TFH}, what we learn from Table 3 in addition is that there are no other combinations
of $(k_e,k_\mu,s_e,s_\mu)$  that is structurally sound for $(m,p)=(4,7)$, at least not for groups of order $<512$.

$$\ba{|c|c|c|c|c|c|c|c|}\hline
{\rm SmallGroup}&(s_e, s_\mu)&a=(a_R,a_I)&(i,j,k,{\rm IR}_3)&c_2&c_3&c_{5}&{\rm IR}_3\\ \hline
[84,11]&-&-&-&3&-&4,9&4\!\!-\!\!12\\ 
\hline
[168,42]&(1,2)&\o_7(1+\o_7+\o_7^3)&(2,4,5,3),(2,4,6,2)&2&4&5,6&2,3\\ 
&&=(-0.5,1.323)&&&&&\\
&(3,5)&\o_7^3(1+\o_7^2+\o_7^3)&(2,4,5,2),(2,4,6,3)&&&&\\
&&=(-0.5,-1.323)&&&&&\\
 \hline
[336,57]&-&-&-&5&3,7,9,10&4,20& 4\!\!-\!\!40\\
\hline\ea$$
\bc Table 4. Allowed groups and allowed parameters for $(m,p)=(4,7)$\ec

\subsection{$[600,179]=\Delta(600)$}
We can also use this technique to find all the allowed parameters $(m,k_e,k_\mu,p,s_2,s_\mu)$ for a given group. As an 
illustration let us consider
$\Delta(600)$ with $(m,k_e,k_p)=(3,0,1)$. The allowed values of $(p,s_e,s_\mu)$ and $a$, together with
 the equivalent mixing vectors $|v|$ calculated from \eq{mv3}, are listed in Table 5. 

Since the classes of $\Delta(600)$ have orders 1-5, 10, 20, we will consider $p=3,4,5,10,20$.

$$\ba{|c|c|c|c|}\hline
p&(s_e,s_\mu)&a=(a_R,a_I)&|v|\\ \hline
3&(0,1)&0=(0,0)&(.577,.577,.577)\\ \hline
4&(0,1)&1=(1,0)&(.816,.408,.408)\\
&(0,2)&-1=(-1,0)&(.000,.707,.707)\\ \hline
5&-&-&-\\ \hline
10&(0,5)&-1=(-1,0)&(.000,.707,.707)\\ 
&(1,2)&-\o_5^3=(.809,.588)&(.777, .607,  .170)\\
&(3,6)&-\o_5^4=(-.309,.951)&(.480, .812, .332)\\
&(3,8)&-\o_5^2=(.809,-.588)&(.777, .170, .607)\\
&(4,9)&-\o_5=(-.309,-.951)&(.480,  .332, .812)\\ \hline
20&(0,5)&1=(1,0)&(.816,.408,.408)\\
&(1,8)&\o_5^2=(-.809, .588)&(.252, .800, .546)\\
&(2,4)&-\o_5^3=(.809,.588)&(.777, .607,  .170)\\
&(2,6)&-\o_5^4=(-.309,.951)&(.480, .812, .332)\\
&(3,4)&\o_5=(.309,.951)&(.661, .746, .086)\\
&(4,13)&\o_5=(.309,.951)&(.661, .746, .086)\\
&(6,16)&-\o_5^2=(.809,-.588)&(.777, .170, .607)\\
&(7,16)&\o_5^4=(.309,-.951)&(.661, .086, .746)\\
&(8,18)&-\o_5=(-.309,-.951)&(.480,  .332, .812)\\
&(9,12)&\o_5^3(-.809, -.588)&(.252, .546, .800)\\ \hline
\ea$$
\bc Table 5. Allowed $(p,s_e,s_\mu,a)$ and mixing vector $|v|$ for $\Delta(600)$ with $(m,k_e,k_\mu)=(0,1,2)$\ec

A neutrino mixing matrix with the third column given by the mixing
vector (.170, .607, .177) and the second column given by the mixing vector  (.577, .577, .577), both appearing in Table 5,
is in good agreement with the measured neutrino mixing parameters \cite{Hu, Lam1, HLL, Lam2}.

\appendix
\section{GAP Programs}
\subsection{Usage}
Two programs written in the GAP language \cite{GAP} are given in this Appendix. 

The main program is the function {\bf gpsearch(m,p,g,LS)}. It 
tests whether the structure constant $\g_{ijk}$ (see \eq{stcon}) for the group $G=$SmallGroup($g$) is nonzero for some $i\in c_2,\ j\in c_m$, and $k\in c_p$
(see Sec.~IIIA). If not, it returns `structurally forbidden'. If so, it gives different printouts
depending on whether the short form (LS=1) or the long form (LS=2) is chosen.

If LS=1, all the allowed $[k_e,k_\mu,s_e,s_\mu,a]$ (see Sec.~II) are printed out. If LS=2,  then the printout includes
the lists $c_2, c_m, c_p$ and the list IR$_3$ of 3-dimensional irreducible representations of $G$.
In addition, also a list of $[[i,j,k,n], [k_e,k_\mu,k_\tau], [s_e,s_\mu,s_\tau],a]$  for which $\g_{ijk}>0$.
The allowed parameters $(k_e,k_\mu,k_\tau, s_e,s_\mu,s_\tau,a)$ depend on the irreducible representation $n\in$ IR$_3$.

The program stops when one of the following happens. Either the order of $G$ is not a multiple of the least common multiple
of (2,$m$,$p$), or there is no 3-dimensional irreducible representation in $G$, or one of $c_2, c_m, c_p$ is empty.

The second function {\bf gpsu3(m,p)} runs gpsearch(m,p,g,1) through all g which is a
finite $SU(3)$ subgroup of even order less
than 512.
\subsection{gpsearch}
\subsubsection{\underline{Explanations}}
\begin{itemize}
\item lines 04-10 defines a sub-function which converts a list L of cyclotomic numbers E(n)$^k:=\o_n^k$ into a list of 
the exponents $k$.
\item the lists of $c_2, c_m, c_p$, and IR$_3$ are computed in lines 24-26 and line 20.
\item the lists dp2, dm2, dp2 in line 35 are sublists of $c_2, c_m, c_p$ with determinant=+1.
\item the do loop between lines 28 and 59 runs through all $n\in$IR$_3$.
\item the three do loops between lines 40 and 50 run through all $i\in$ dp2, $j\in$ dm2, and $k\in$ dp2.
\item $\g_{ijk}$ is calculated in line 41.
\item the eigenvalues of $T$ are calculated in line 43, and converted to the index form $[k_e,k_\mu,k_\tau]$ in line 47.
\item the eigenvalues of $W_U$ are calculated in line 48, and converted to the index form $[s_e,s_\mu,s_\tau]$ in line 52. 
\item $a=$Tr($W_U$) is calculated in line 50.
\item the rest of the lines prepares for the outputs, depending on whether LS=1 or LS=2.
\end{itemize}

\subsubsection{\underline{The Program}}
{\small
\ni 01\hs gpsearch:=function(m,p,g,LS)  \nl
\ni 02\hs  local convert,out,class,n,G,dir,ir3,order,p2,pm,pp,iir3,d2,dm,dp,dp2,dpm,dpp,tbl,ip2,\nl
\ni 03\hs\hs\hs ipm,ipp,evm,evp,a,kevm,sevp,out1,out2,out3;\nl
\# -----------------------------------------------------------------\nl
\ni 04\hs\hs
      convert:=function(L,n)\hs  \# L is a list, with L[i]=E(n)\^k[i]. Outputs k[i]\nl
\ni 05\hs\hs\hs         local i,j,k,nl;\nl
\ni 06\hs\hs\hs         nl:=Number(L); k:=List([1..nl],x$\to$0); \nl
\ni 07 \hs\hs\hs for i in [1..nl] do\quad 
         for j in [0..n-1]do\nl
\ni 08\hs\hs\hs\hs   if E(n)\^j=L[i] then k[i]:=j; break; fi; \nl
\ni 09 \hs\hs\hs od; od; return(k);\nl
\ni 10\hs\hs end;;\nl
 \# ------------------------------------------------------------------\nl
\ni 11\hs if LS$<>$1 and LS$<>$2 then \nl
\ni 12 \hs\hs return("the LS argument has to be either 1 (short) or 2 (long)"); fi;\nl
\ni 13 \hs out:=[\ ]; \nl
\ni 14 \hs  if not IsInt(g[1]/LcmInt(2,LcmInt(m,p))) then return("group has wrong order"); fi;\nl   
\ni 15\hs\hs      \# stops the program if lcm(2,m,p) does not divide the order of the group\nl
\ni 16\hs       G:=SmallGroup(g);\quad   \# G is the group to be searched\nl  
\ni 17\hs   class:=ConjugacyClasses(G); \quad \# a list of conjugacy classes of G\nl  
 \ni 18\hs  dir:=List([1..Number(class)], i$\to$Trace(Identity(G)\^IrreducibleRepresentations(G)[i]));\nl
\ni 19\hs\hs      \# list of dimensions of irreducible representations of G\nl
\ni 20\hs   ir3:=Positions(dir,3);\quad  \# a list of 3-dim irreducible representations of G\nl
\ni 21\hs  if ir3=[\ ] then return("this group has no 3-dim irred rep"); fi; \nl 
\ni 22\hs\hs     \# stops the program if there are no 3-dimensional irr rep\nl
\ni 23\hs    order:=List(class, x$\to$Order(Representative(x)));\quad  \# list of  class orders\nl 
\ni 24\hs    p2:=Positions(order,2);\quad \# a list of order-2 classes\nl
\ni 25\hs    pm:=Positions(order,m);\quad \# a list of order-m classes\nl
 \ni 26\hs     pp:=Positions(order,p);\quad \# a list of order-p classes\nl
\ni 27\hs  if p2=[\ ] or pm=[\ ] or pp=[\ ] then return("this group contains no such orders"); fi;\nl
 \ni 28\hs for iir3 in ir3 do\nl
 \ni 29\hs\hs   d2:=Positions(List(p2,x$\to$Determinant(Representative(\nl
\ni 30\hs\hs\hs class[x])\^IrreducibleRepresentations(G)[iir3])),1);\nl
\ni 31\hs\hs     dm:=Positions(List(pm,$\to$Determinant(Representative(\nl
\ni 32\hs\hs\hs class[x])\^IrreducibleRepresentations(G)[iir3])),1);\nl
\ni 33\hs\hs     dp:=Positions(List(pp,x$\to$Determinant(\nl
\ni 34\hs\hs\hs Representative(class[x])\^IrreducibleRepresentations(G)[iir3])),1);\nl
\ni 35\hs\hs     dp2:=List(d2,i$\to$p2[i]); dpm:=List(dm,i$\to$pm[i]); dpp:=List(dp,i$\to$pp[i]);\nl
 \ni 36\hs\hs\hs     \# for each 3-dimensional irreducible representation iir3, \nl
\ni 37\hs\hs\hs \# dp2, dpm, dpp are sublists of lists p2, pm, pp with determinant=1\nl
\ni 38\hs\hs     if d2=[\ ] or dm=[\ ] or dp=[\ ] then continue; fi; \nl
\ni 39\hs\hs     tbl:=CharacterTable(G);  \# character table of G used to calculate structure constant\nl
\ni 40 \hs\hs for ip2 in dp2 do\quad   for ipm in dpm do\quad   for ipp in dpp do\nl
\ni 41\hs\hs\hs if ClassMultiplicationCoefficient(tbl, ip2,ipm,ipp)=0  
     then continue;\nl
\ni 42\hs\hs\hs\hs    \# select out non-zero structure constants gamma[ip2,ipm,ipp]\nl
 \ni 43\hs\hs\hs evm:=Eigenvalues(CF(m),Representative(\nl
\ni 44\hs\hs\hs\hs class[ipm])\^IrreducibleRepresentations(G)[iir3]); \nl
\ni 45\hs\hs\hs\hs\hs      \# eigenvalues of T in that irreducible rep\nl
\ni 46\hs\hs\hs if Number(evm)$<$3 then continue; fi;\quad  \# proceed only if T is nondegenerate\nl
 \ni 47\hs\hs\hs      kevm:=convert(evm, m); 
        \# convert eigenvalues of T into exponents [ke, km, kt]\nl
\ni 48\hs\hs\hs       evp:=Eigenvalues(CF(p),Representative(\nl
\ni 49\hs\hs\hs\hs class[ipp])\^IrreducibleRepresentations(G)[iir3]); \nl
\ni 50\hs\hs\hs       a:=Trace(Representative(class[ipp])\^IrreducibleRepresentations(G)[iir3]);\nl
\ni 51\hs\hs\hs\hs        \# distinct eigenvalues and trace of W\_U\nl
\ni 52\hs\hs\hs       sevp:=convert(evp,p);
       \# convert eigenvalues of W\_U into exponents [se, sm, st]\nl
 \ni 52\hs\hs\hs out1:=Concatenation([kevm],[sevp],[a]);\quad \# list [kevm, sevp, a]\nl
\ni 53\hs\hs\hs out2:=Concatenation([[ip2,ipm,ipp,iir3]],out1); \# list [[ip2,ipm,ipp,iir3],[kevm,sevp,a]]\nl
 \ni 54\hs\hs\hs out3:=[kevm[1],kevm[2],sevp[1],sevp[2],a];\quad\# list [ke,km,se,sm,a]\nl
  \ni 55\hs\hs\hs if LS=2 then out:=Concatenation(out, [out2]);\nl
\ni 56\hs\hs\hs\hs \# string lists out2 together into a big list out \nl
\ni 57\hs\hs\hs elif LS=1 then out:=Concatenation(out, [out3]); fi;\nl
\ni 58\hs\hs\hs\hs \# string lists out3 together into a big list out \nl
\ni 59 \hs od;od;od;od;\quad \# close iir2, ip2, ipm, ipp loops\nl
\ni 60 \hs if out=[\ ] then return("structurally forbidden$\backslash$n"); fi;\nl
\ni 61 \hs if LS=2 then Print("$\backslash$nc2=",p2,  "  cm=",pm,  "  cp=",pp,"  IR3=",ir3,"$\backslash$n"); fi;\nl
 \ni 62\hs if LS=2 then Print("[ip2,ipm,ipp,iir3], [ke,km,kt], [se,sm,st],a$\backslash$n"); fi;  \nl
\ni 63\hs return(out); end;;
} 
\subsection{gpsu3}

{\small
\ni 01\hs gpsu3:=function(m,p)\quad  \# search SU(3) finite subgroups of order $<$512\nl
 \ni 02\hs local su3fil, g;\nl
\ni 03\hs  su3fil:=[ [ 12, 3 ], [ 24, 12 ], [ 48, 3 ], [ 54, 8 ], [ 60, 5 ], [ 84, 11 ], 
  [ 96, 64 ], \nl
\ni 04\hskip2cm [ 108, 15 ], [ 108, 22 ], [ 150, 5 ], [ 156, 14 ], [ 162, 14 ], 
  [ 168, 42 ], \nl
\ni 05\hskip2cm [ 192, 3 ], [ 216, 88 ], [ 216, 95 ], [ 228, 11 ], [ 294, 7 ], 
  [ 300, 43 ],\nl
\ni 06\hskip2cm  [ 324, 50 ], [ 336, 57 ], [ 372, 11 ], [ 384, 568 ], 
  [ 432, 103 ],\nl
\ni 07\hskip2cm [ 444, 14 ], [ 486, 61 ] ];;\quad  \# even-order finite subgroups of SU(3)\nl
\ni 08\hs   for g in su3fil do  if not IsInt(g[1]/LcmInt(2,LcmInt(m,p))) then continue; fi;\nl
\ni 09\hs  Print(gpsearch(m,p,g,1),"$\backslash$n$\backslash$n"); od; end;;

}

\edoc
\begin{thebibliography}{9}
\bibitem{HS} D. Hernandez, A. Yu. Smirnov, Phys.~Rev.~D 86 (2012) 053014 [arXiv:1204.0445];
arXiv:1212.2149.
\bibitem{Hu} B. Hu, arXiv:1212.2819.
\bibitem{Lam1}  C.S. Lam, arXiv:1208.5527, to appear in Phys.~Rec,~D.
\bibitem{HLL} M. Holthausen, K.S. Lim, M. Lindner, arXiv:1212.2411.
\bibitem{TFH} R. de Adelhart Toorop, F. Feruglio and C. Hagedorn, Nucl.~Phys. B 858 (2012)
437 [arXiv:1112.1340].
\bibitem{Lam2} C.S. Lam, arXiv:1301.1736.
\bibitem{GAP} www.gap-system.org.
\bibitem{Ludl} P.O. Ludl, J.~Phys.~A 43 (2010) 395204; 44 (2011) 139501(E).
\end{thebibliography}
